\documentclass[pdflatex,aps,prd,nofootinbib,showpacs,superscriptaddress,floatfix, twocolumn]{revtex4-2}

\usepackage{bm,amsmath,amssymb}
\usepackage{graphicx}
\usepackage{subfigure}
\usepackage{color}
\usepackage{soul}
\usepackage{hyperref}
\usepackage{multirow}
\usepackage{lineno}
\usepackage{tabularx}
\usepackage{float}

\setulcolor{blue}
\setstcolor{red}
\sethlcolor{yellow}
\definecolor{brown}{rgb}{0.5, 0.25, 0.0}

\newlength{\wdth}

\newcommand \beq {\begin{equation}}
\newcommand \eeq {\end{equation}}
\newcommand \beqa {\begin{eqnarray}}
\newcommand \eeqa {\end{eqnarray}}

\newcolumntype{P}[1]{>{\centering\arraybackslash}p{#1}}

\def\lsim{\raise0.3ex\hbox{$<$\kern-0.75em\raise-1.1ex\hbox{$\sim$}}}
\def\gsim{\raise0.3ex\hbox{$>$\kern-0.75em\raise-1.1ex\hbox{$\sim$}}}

\begin{document}

\title{Finite-size scaling analysis of three dimensional \texorpdfstring{\boldmath{$Z(2)$}}{Z2} and \texorpdfstring{\boldmath{$O(2)$}}{O2} spin models with non-vanishing symmetry breaking
parameter} 

\author{Jishnu Goswami}
\email{jishnu@physik.uni-bielefeld.de}
\affiliation{Fakult\"at f\"ur Physik, Universit\"at Bielefeld, D-33615 Bielefeld,
 Germany}
\author{Frithjof Karsch}
\affiliation{Fakult\"at f\"ur Physik, Universit\"at Bielefeld, D-33615 Bielefeld,
Germany}
\author{Sabarnya Mitra}
\affiliation{Fakult\"at f\"ur Physik, Universit\"at Bielefeld, D-33615 Bielefeld,
Germany}

\begin{abstract}
We present results from a detailed finite-size scaling analysis of 
the $3$-$d$,  $Z(2)$ and $O(2)$ spin models in an external field $H$. Using high statistics 
Monte Carlo data, we obtain the leading finite-size 
scaling correction to the infinite volume scaling functions. 
We show that these corrections are 
proportional to $\widetilde{z}_L^2=1/(H L^{3/(1+1/\delta))})^{2}$. 
This provides the parametric form for finite-size corrections to bulk thermodynamic observables as well as the pseudo-critical temperatures
determined at non-vanishing $H$. In particular, it
allows to eliminate systematic errors in the
analysis of the chiral phase transition temperature in (2+1)-flavor QCD, that arose 
from the previously not well-controlled ansatz for infinite volume extrapolations. We also establish the validity range of this leading order correction and point out that there are significant differences between the $3$-$d$, $O(2)$ and $Z(2)$
universality classes.
\end{abstract}


\maketitle

\section{Introduction}
The renormalization group (RG) approach to critical behavior formulated by 
K. Wilson \cite{Wilson:1971bg,Wilson:1971dh,Domb:1976bk} relates properties
in the vicinity of a critical point to the scale invariance of
renormalization group transformations in the vicinity of a fixed point. It also
provides a natural framework
for the discussion of universal and non-universal aspects of finite-size 
scaling (FSS) in the vicinity of a second order phase transition
\cite{Fisher:1972zza,Binder:1981sa,Privman:1984zz}. This finds widespread
application in the study of the phase structure of various statistical and
field theoretic models of interest in condensed matter, nuclear and high
energy physics \cite{Zinn-Justin:2002ecy,Cardy:1996xt}. 

The RG approach provides an explicit parametrization of the singular 
part of the free energy density ($f_s$) of a statistical model in terms of universal 
parameters (critical exponents $y_t$, $y_h$)
characterizing the universality class of a $d$-dimensional model and 
non-universal scale parameters ($T_c,t_0, H_0 ,L_0$)
which set the scale for
deviations of temperature ($T$),
symmetry breaking parameter ($H$)
and spatial size ($L$)
from the critical point.
Scale invariance near a RG fixed point is encoded in
the appearance of an arbitrary scale parameter $b$,
\begin{equation}
f_s(t,h,\ell)= b^{-d}f_s(b^{y_t}t,b^{y_h}h,b\,\ell) \; ,\
        \label{RG}
\end{equation}
with reduced temperature $t=(T-T_c)/(T_c t_0)$, symmetry breaking parameter 
$h=H/H_0$ and volume parameter $\ell=L_0/L$. The RG scaling exponents $y_t$ and $y_h$
are related to the critical exponents of a $d$-dimensional universality class,

\begin{equation}
y_t\;=\;1/\nu\;,\; y_h\;= \;1/\nu_c = \beta\delta/\nu\; ,
\end{equation}
with $\nu$ being related to $\beta$ and $\delta$ through a 
hyper-scaling relation $d\nu=\beta (1+\delta)$, {\it i.e.} $\nu_c=(1+1/\delta)/d$.

The parameter $b$
can conveniently be chosen such that one of the scaled arguments in Eq.~\ref{RG}
is set to unity. Standard choices are 
(i) $b=1/\ell$ or (ii) $b=1/h^{1/y_h}$. This leads to two scaling functions,
$Y_f$ and $f_f$, depending only on two arguments,
\begin{eqnarray}
f_s(t,h,\ell)\hspace{-0.1cm}&=& \hspace{-0.1cm}
	\begin{cases}
		\ell^{d}\ Y_f(t\,\ell^{-y_t},h\,\ell^{-y_h}) &\hspace{-0.3cm}, b=\ell^{-1} \; , \label{RGLYf}\\
		h^{d/y_h}f_f(t\,h^{-y_t/y_h},\ell\,h^{-1/y_h}) &\hspace{-0.3cm}, b=h^{-1/y_h}~. \label{RGhfs}
	\end{cases}
\end{eqnarray}
Obviously, the scaling functions $Y_f$ and $f_f$ are related to each other,
e.g. $f_f(z,z_L) \equiv z_L^{-d}\, Y_f(z z_L^{-y_t},z_L^{-y_h})$, with 
\begin{equation}
z=\frac{t}{h^{y_t/y_h}} \;\; ,\;\; z_L=\frac{\ell}{h^{1/y_h}} \; .
\label{zzL}
\end{equation}
The choice $b=1/\ell$ puts the volume dependence directly in front of the scaling function $Y_f$ and as such is
convenient to extract non-universal as well as universal 
scale parameters from a FSS analysis of thermodynamic observables at vanishing external field $H$
\cite{Binder:1981sa,Cardy:1996xt,Campostrini:2000iw}.
The choice $b=1/h^{1/y_h}$, on the other hand, puts the dependence on $H$ in front of the scaling function $f_f$. As such it is favorable in studies of finite-size corrections to
the infinite volume limit of thermodynamic observables at fixed, non-vanishing values of
$H$~\cite{Engels:2014bra}.
This situation is frequently met in studies of field theories when aiming at a 
comparison of analytic calculations, that are usually performed for an infinite system, with numerical studies on finite, discrete space-time lattices.

Finite-size corrections to the infinite volume scaling function $f_f(z)\equiv f_f(z,0)$ of $3$-$d$ spin
models in the $Z(2)$ and $O(N)$ universality classes have been calculated \cite{Engels:2014bra,Karsch:2023pga} and 
parametrized for intermediate values of $z_L$ and also the large volume
behavior of $Y_f$ has been analyzed \cite{Springer:2015kxa}. However, no systematic investigation
of the leading order finite-size
corrections to $f_f(z)$ 
has been performed so far.
In particular, we stress that the finite-size scaling variable $z_L$, appearing in $f_f(z,z_L)$, parametrizes the finite-size corrections to $f_f(z)$ only to the extent that it fixes the parametric relation between $\ell$ and $h$. This, however, 
does not at all mean that $z_L$ is an
appropriate expansion parameter of 
$f_f(z,z_L)$ close to the infinite-volume limit, $z_L\rightarrow 0$. 
The scaling variable $\widetilde{z}_L=z_L^{\beta\delta/\nu}$, appearing in $Y_f$ may equally well
control the large volume behavior.

The purpose of this work is therefore, to determine the leading order finite-size correction
to the scaling function $f_f(z,z_L)$
in the 
$3$-$d$, $Z(2)$ and $O(2)$
universality classes. We will show in the following that this is
given by\footnote{We keep
$z_L$ as argument in the scaling functions, as is commonly done. However we write
expansions of the scaling functions in terms of $\widetilde{z}_L$ as we show
in this work that this is the
natural expansion parameter.
}
\begin{equation}
    f_f(z,z_L) = f_f(z,0) + a_f(z) \ \widetilde{z}_L^2 \;,\;
    {\rm with} \;\; \widetilde{z}_L =z_L^{1/\nu_c} \; .
\end{equation}
\begin{figure*}[t]
\includegraphics[width=0.48\textwidth]{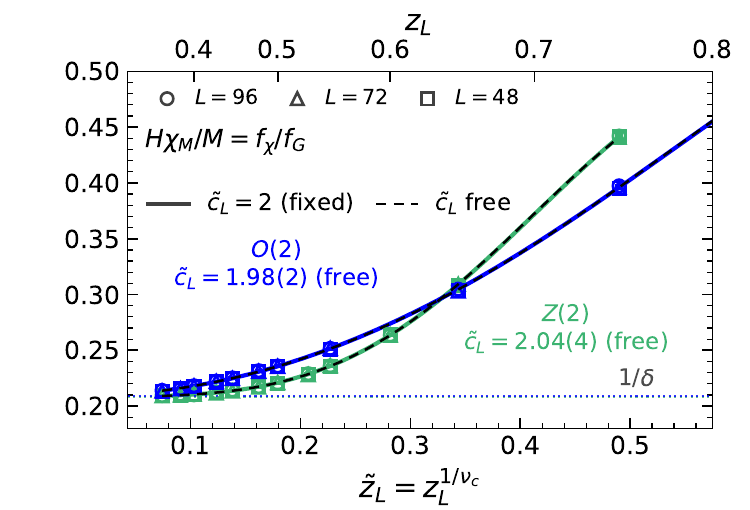}
\includegraphics[width=0.48\textwidth]{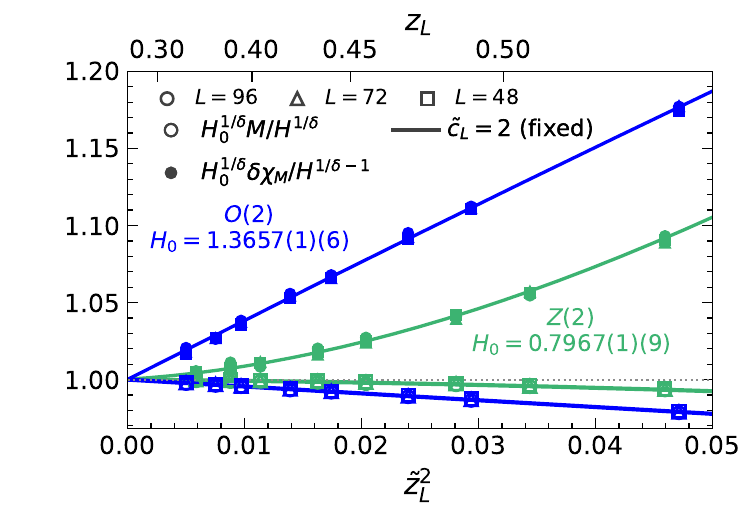}
\caption{({\it Left}:) The ratio $H\chi_M(T_c,H,L)/M(T_c,H,L)$
versus the scaling variable $\widetilde{z}_L$
for the $3$-$d$, $Z(2)$ and $O(2)$ spin
models on cubic lattices 
of size $L^3$
with $L=48, 72$ and $96$.
({\it Right}:) The scaled 
order parameter $M/h^{1/\delta}$ and scaled susceptibility $H_0\delta \chi_M/h^{1/\delta -1}$ versus
$\widetilde{z}_{L}^2$ for both $Z(2)$ and $O(2)$ model.
While in the $O(2)$ universality class the leading order term provides 
a good description for $z_L\le 0.55$, a significant
influence of higher order terms in the expansion 
are seen for the $Z(2)$ 
universality class above 
$z_L\simeq 0.4 $
(see text).
Note that the conversion from the $\widetilde{z}_L$ scale shown on the lower x-axis to the $z_L$ scale shown on the upper x-axis is done for the $O(2)$ case. Differences for 
the same in the $Z(2)$ case are however, comparatively minor.
}
\label{fig:FSSsmall}
\end{figure*}

\section{Leading order corrections to scaling functions}
We want to calculate the leading order FSS correction to the free energy scaling
function $f_f(z,z_L)$, which controls the asymptotic, large volume behavior of
bulk thermodynamic observables in the $3$-$d$, $Z(2)$ and $O(2)$ universality
classes.
In particular, we will
establish the parametric form of the FSS dependence of the singular part, $M(T,H,L)_{sing}$,
of the order parameter $M(T,H,L)$, 
\begin{eqnarray}
M(T,H,L)_{sing} &=&-\frac{\partial f_f(z,z_L)}{\partial H}\equiv h^{1/\delta}f_G(z,z_L)\; ,
\label{eq:magn}
\end{eqnarray}
and the corresponding singular 
contribution, $\chi(T,H,L)_{sing}$, to the order parameter susceptibility $\chi_M$, 
\begin{eqnarray}
\chi_M(T,H,L)_{sing} &=& \frac{\partial M(T,H,L)_{sing}}{\partial H} \nonumber \\
&\equiv& h^{1/\delta-1}H_0^{-1}f_\chi(z,z_L) \; .
\label{eq:chin}
\end{eqnarray}
The corresponding scaling functions are given by
\begin{eqnarray}
        f_{G} (z,z_L) &=&
-\left(1+\frac{1}{\delta}\right) f_f(z,z_L)
+\frac{z}{\beta\delta}\frac{\partial f_f(z,z_L)}{\partial z} \nonumber \\
	&&+\frac{\nu z_L}{\beta\delta}\frac{\partial f_f(z,z_L)}{\partial z_L} \ ,
\label{fgff}\\
        f_{\chi}(z,z_L) &=& \frac{1}{\delta} f_{G}(z,z_L)
- \frac{z}{\beta\delta} \frac{\partial f_{G} (z,z_L)}{\partial z} \nonumber \\
&&   -\frac{\nu z_L}{\beta\delta} \frac{\partial f_{G} (z,z_L)}{\partial z_L}\; .
        \label{fchis}
\end{eqnarray}
We first analyze the leading FSS correction at $z=0$, using the ansatz,
\begin{equation}
	f_f(0,z_L)=-\frac{\delta}{1+\delta} + a_f \,z_L^{c_L} 
	\equiv -\frac{\delta}{1+\delta} + a_f \,\tilde{z}_L^{\,\widetilde{c}_L}
	\; ,
	\label{ff0}
\end{equation}
with $c_L=\widetilde{c}_L/\nu_c$. This implies
\begin{eqnarray}
	\hspace{-.8cm} f_G(0,z_L) &=& 1+a_G \tilde{z}_L^{\widetilde{c}_L} \; ,\; 
	a_G = a_f\bigg( -\frac{\delta + 1}{\delta} +\widetilde{c}_L \biggr)
 \; ,\label{fGasym}\\
	\hspace{-.8cm} f_\chi(0,z_L) &=& \frac{1}{\delta} +a_\chi \tilde{z}_L^{\widetilde{c}_L} \; ,\; a_\chi=a_G\biggl( \frac{1}{\delta}-\widetilde{c}_L \biggr) \; . \label{fchiasym}
\end{eqnarray}
For $z_L>0$, we thus may determine the coefficient $\widetilde{c}_L$ from the ratio,
\begin{equation}
\left(\frac{H\chi_M}{M}\right)_{sing} = \frac{f_\chi(0,z_L)}{f_G(0,z_L)} =\frac{1}{\delta} -a_G \widetilde{c}_L \widetilde{z}_L^{\widetilde{c}_L} \; ,
\label{HchiM}
\end{equation}
or directly from
\begin{equation}
        \frac{f_\chi(0,z_L)-1/\delta}{f_G(0,z_L)-1} =
	\frac{1}{\delta}-\widetilde{c}_L 
\; .
	\label{eq:ratio}
\end{equation}

\subsection{Numerical FSS analysis at \texorpdfstring{$z=0$}{z=0}}
Previous studies of the FSS functions of
the $3$-$d$, $Z(2)$ and $O(2)$
models lead to a parametrization of the
scaling function 
in terms of a double series in $z$ and
$z_L$ \cite{Engels:2014bra,Karsch:2023pga}. 
The parameters of this series have been obtained from a global, joint fit to data for $M$ and $\chi_M$. 
Naturally, such a fit is not meant to provide a parametrically correct description of the scaling functions in terms of a Taylor series. 
In particular,
although calculations have also been performed for $z_L\le 0.4$, the statistics in this region was not
sufficient for a systematic
analysis of the small $z_L$ regime. 

For our analysis of the small $z_L$ region, we performed new calculations 
for the 
$3$-$d$, $Z(2)$ and $O(2)$ spin models.
These calculations have been performed in the entire range $z_L\in [0.3:1.1]$.
However, the focus has been 
put on the region
$z_L\le 0.55$. 
The simulations have been performed for the 
improved $Z(2)$ and $O(2)$ spin models with the Hamiltonian,
\begin{eqnarray}
 \mathcal{H} &=& -J \sum_{\left\langle x,y \right\rangle} \Phi_x \Phi_y + \sum_x \left[ \Phi^2_x + \lambda \left(  \Phi^2_x -1 \right)^2  \right]
 \nonumber \\
   &&- H \sum_x \Phi_{x,1} \; ,
    \label{Hamiltonian}
\end{eqnarray}
with real valued $\Phi_x \equiv (\Phi_{x,1}\;,\;\Phi_{x,2})$
for the $O(2)$ model and $\Phi_x\equiv \Phi_{x,1}$ for $Z(2)$. A term proportional to the coupling $\lambda$ has
been introduced to suppress contributions from
sub-leading corrections-to-scaling
in universal scaling relations \cite{Wegner:1972my}. We use 
 $\lambda=1.1$ for the $Z(2)$ model \cite{Hasenbusch:1999mw}  and $\lambda=2.1$ for the $O(2)$ model 
\cite{Hasenbusch:1999cc}.  
Further details of our computational setup are given in \cite{Karsch:2023pga} and in the appendix~\ref{app:fit_details}.
As we want to firmly establish the parametric dependence of FSS scaling 
functions in the limit $z_L\rightarrow 0$ utilizing
high statistics calculations at $z=0$, {\it i.e.} at the 
phase transition temperature $T_c=1/J_c$, we need for our 
calculations, high precision knowledge of the critical coupling $J_c$. 
In the context of our FSS study, we re-calculated the 
inverse critical temperature from the 
unique intersection point of the rescaled order parameter $M_{sing}/H^{1/\delta}$
at $J_c$. For small $z_L$,
this intersection point is
volume independent. Furthermore, the value 
of $M/H^{1/\delta}$ at $J_c$
provides a new determination
of the non-universal scale parameter $H_0$. In addition, we also re-calculated the non-universal scale parameter $L_0$, which sets the scale for $z_L$.
Details of this calculation are given in the appendix. While our new result for $H_0$, $L_0$ agrees within errors with those given in \cite{Karsch:2023pga},
we find for the inverse critical temperature of the $3$-$d$, 
$Z(2)$ model $J_c=0.3750966(4)$
and 
$O(2)$ spin model
$J_c=0.5091463(10)$. 
The latter is about
3$\sigma$ smaller than the
value quoted in \cite{Campostrini:2000iw}.

We calculated the order parameter $M$ and the susceptibility $\chi_M$ on
lattices of size $L=48,\ 72$ and $96$ at the critical temperature $T_c=1/J_c$ for the $3$-$d$, $Z(2)$ and $O(2)$ models. 
Results for the ratio $H\chi_M/M$ plotted versus
the scaling variable 
$\widetilde{z}_{L,b}=1/(H L^{1/\nu_c})$
are shown in Fig.~\ref{fig:FSSsmall}~(left).
No significant deviations from universal FSS behavior are visible in this figure,
suggesting that contributions from regular terms are negligible.
On fitting the data for $H\chi_M/M$ to the scaling 
ansatz 
\begin{eqnarray}
	\frac{f_\chi(0,z_L)}{f_G(0,z_L)} &=& \frac{1}{\delta}+\sum_{n=1}^{n_{max}} r_{2n} \widetilde{z}_L^{\,n\,\widetilde{c}_L} \; 
	\label{eq:eqntext},
\end{eqnarray}
with $n_{max}=1$-$3$,
we find for $O(2)$ and $Z(2)$,
\begin{equation}
    \widetilde{c}_L^{O(2)} = 1.98 (2)
    \;\; , \;\; 
  \widetilde{c}_L^{Z(2)} = 2.04(4)
\;\; .   
\end{equation}
We thus conclude that FSS corrections to the free energy density, as well as the 
order parameter and its susceptibilities are indeed 
controlled in leading order by
$\widetilde{z}_L^2\sim 1/(H L^{1/\nu_c})^2$. This is also apparent from Fig.~\ref{fig:FSSsmall}~(right), where we show the scaled order parameter $M/h^{1/\delta}$ and the 
scaled susceptibility $\delta H_0\chi_M/h^{1/\delta-1}$
as function of $\widetilde{z}_{L,b}^2$ in 
a small $\widetilde{z}_L^2$
interval.
The lines show results from the leading ($n_{max}=1$) and next-to-leading  ($n_{max}=2$) order 
expansions in terms of $\widetilde{z}_L^2$.
This confirms
that $\widetilde{z}_L^2$ indeed
is the appropriate expansion parameter
for the leading order finite-size corrections to the FSS functions of the free energy in the $3$-$d$, $Z(2)$ and $O(2)$ 
universality classes. Furthermore, it shows the 
importance of next-to-leading order corrections particularly in the $Z(2)$ universality class.
The first two expansion coefficients, $a_{G,2}$ and
$a_{G,4}$ for the finite-size correction in $f_G(z,z_L)$ 
are easily obtained from
$r_2$ and $r_4$:
$a_{G,2}=-r_2/2$ and $a_{G,4}=r_2^2/8-r_4/4$.
These coefficients 
are given in Table~\ref{tab:scales}.
Further details on these fits and the resulting fit parameters for $n_{max}>2$, are given in the appendix.

\begin{table}[b]
\begin{center}
\begin{tabular}{|c|c|c|}
\hline
$3$-$d$ universality class &$a_{G,2}$ & $a_{G,4}$\\
\hline
O(2) & -0.4509(20)  & 0.123(9) \\
\hline
Z(2) &  -0.0543(24) & -2.17(3) \\
\hline
\end{tabular}
\end{center}
\caption{Leading order (LO) and next-to-leading order (NLO) coefficients for finite-size
corrections to the order parameter.(See appendix for details.)}
\label{tab:scales}
\end{table}

\begin{figure}[t]
\includegraphics[width=0.48\textwidth]{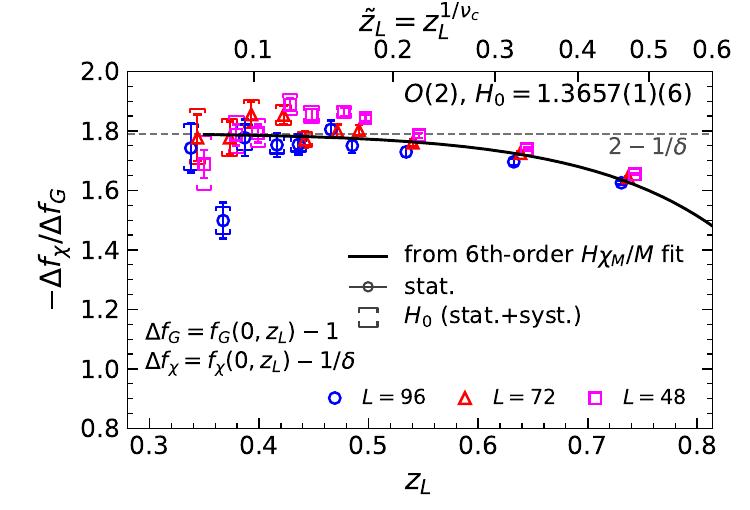}
\caption{The ratio  $-(f_\chi(0,z_L)-1/\delta)/(f_G(0,z_L)-1)$ introduced in Eq.(\ref{eq:ratio})  at $J_c=1/T_c$. Shown are results for 
this ratio obtained from 
calculations for the $3$-$d$,
$O(2)$ spin model.
}
\label{fig:FSSdelta}
\end{figure}

Using the results for the
scaling functions $f_G(z,z_L)$ and $f_\chi(z,z_L)$ obtained from the scaled order parameter $M/h^{1/\delta}$ and the scaled susceptibility $H_0\chi_M/h^{1/\delta}$, we 
also can try to extract the
leading order scaling exponent $\widetilde{c}_L$ from Eq.~\ref{eq:ratio}. This
ratio is shown for the $3$-$d$, $O(2)$ model in 
Fig.~\ref{fig:FSSdelta}.
As can be seen for $z_L\le 0.55$, this ratio is in good
agreement with $\widetilde{c}_L\simeq 2$.

\subsection{Numerical FSS analysis at \texorpdfstring{$z \ne 0$}{zne0}}
\begin{figure}[t]
\begin{center}
\includegraphics[scale=0.484]{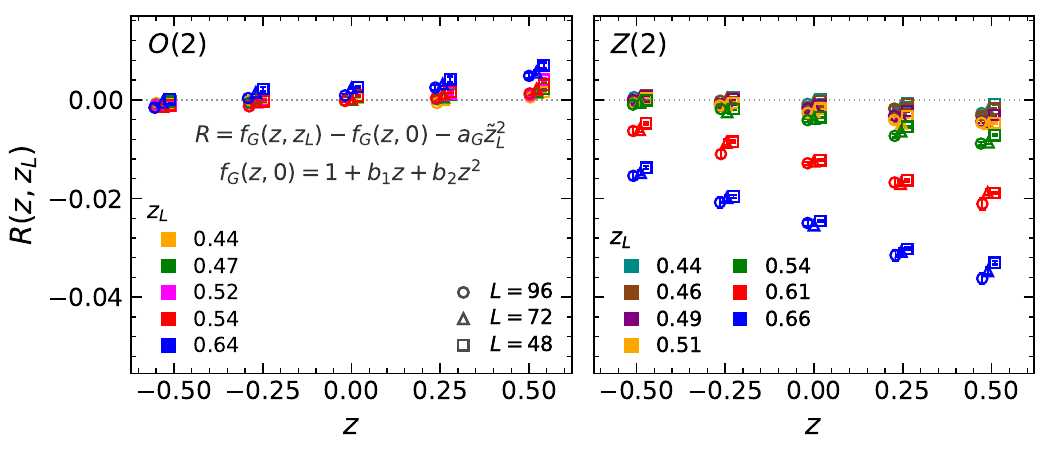}
\end{center}
\caption{Sub-leading corrections to the FSS functions in the $3$-$d$, $O(2)$ ({\it left}) and $Z(2)$ ({\it right}) universality classes versus $z$, for several values of $z_L$, $0.44 \leq z_L \leq 0.66$.
}
\label{fig:FSSz}
\end{figure}

We now extend the above discussion to the case of non-vanishing $z$, corresponding to $T \neq T_c$.
Unlike in the case of the $z_L$-dependence of the scaling functions,
it is well established that the infinite volume scaling functions $f_G(z)$ and $f_\chi(z)$
can be expanded in terms of Taylor series in $z$. In fact, the 
universal expansion coefficients can be calculated in the Widom-Griffith
parametrization. One finds
\cite{Karsch:2023pga}
\begin{equation}
	f_G(z) = 1 + b_1 z + b_2 z^2 +{\cal O}(z^3) \; ,
\end{equation}
with 
$b_1 = -\,0.21866(18)$, $b_2=-\,0.017777(80)$ for the $3$-$d$, $Z(2)$ 
universality class and
$b_1=-\,0.2587(21)$,
$b_2=-\,0.0259(11)$ 
in the case of $O(2)$. This second order expansion 
approximates exact results for $f_G(z)$ to better than 0.1\% for $|z|\le 1$.
In the vicinity of $(z,z_L)=(0,0)$, we thus may compare a series expansion
of $f_G(z,z_L)$ to numerical results for the order parameter calculated
for small $(z,z_L)$. In order to test the importance of corrections
beyond the leading order expansions in $z$ and $z_L$, respectively, we
use for $f_G(z,z_L)$ the ansatz,
\begin{align}
	f_G(z,z_L) &= 1+ b_1 z+ b_2 z^2 + a_{G,2} \widetilde{z}_L^2 + R(z,z_L) \; . 
	\label{Ransatz}
\end{align}
With this, we obtain the ansatz for $f_\chi(z,z_L)$
from Eq.~\ref{fchis}.
In particular, we want to quantify to what extent FSS corrections can
be treated as $z$-independent in the vicinity of $z=0$.
In Fig.~\ref{fig:FSSz}, we show data obtained for $f_G(z,z_L)$
from which we subtracted the leading and next-to-leading order 
infinite volume results for $f_G(z,z_L)$,
as well as the leading finite-volume correction obtained above,
{\it i.e.} we calculate $R(z,z_L)$. 
This is shown in Fig.~\ref{fig:FSSz}.
As can be seen in the left hand panel in the case of $O(2)$ universality class, additional corrections summarized in Eq.~\ref{Ransatz}
in terms of $R(z,z_L)$, are below 0.5\%  for $\widetilde{z}_L\le 0.55$.
This is also the case in
the $Z(2)$ universality class shown in the right hand panel for $\widetilde{z}_L\le 0.50$.
For larger $\widetilde{z}_L$ values
however, higher order corrections rapidly gain in importance.

\section{Pseudo-critical temperatures from FSS analysis}
The universal FSS properties of the  
ratio $H\chi_M/M$ may also be used to define finite-volume estimators $T_\delta(\widetilde{z}_L)$ for
the critical temperature $T_c$. We define $z_\delta(z_L)$ as the point
at which
$\left(H\chi_M/M\right)_{sing}=1/\delta$ .
From the leading and next-to-leading order Taylor
series expansion for the ratio of 
scaling functions, $f_\chi(z,z_L)/f_G(z,z_L)$,
we then obtain the FSS scaling form of
$z_\delta(\widetilde{z}_L)$,
\begin{eqnarray}
z_\delta(\widetilde{z}_L) &=&
k_2 \widetilde{z}_L^2 
+k_4 \widetilde{z}_L^4
\\
k_2=-2 \,a_{G,2}\, \frac{ \beta \delta}{b_1}\; ,\; k_4 &=& -4 \,\frac{\beta \delta}{b_1^3} \,\left(a_{G,4} \,b_1^2  + 2 \,a_{G,2}^2  \,\beta\,\delta \, b_2\right)\,
\; . 
\nonumber
\label{zdelta2}
\end{eqnarray}
\begin{figure}[H]
\includegraphics[width=0.44\textwidth]{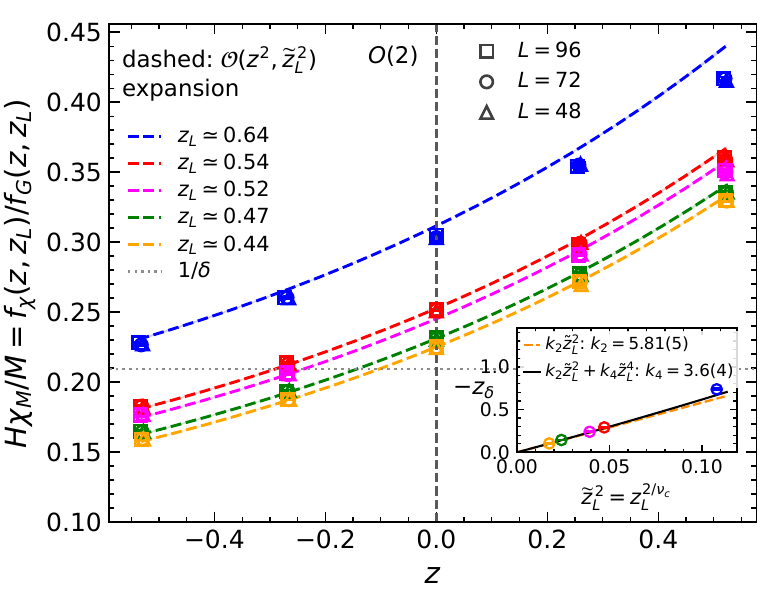}
\caption{The ratio $H\chi_M/M$
versus $z$ for selected $z_L$ values, on our working cubic lattices.
Shown are results for the $3$-$d$, $O(2)$ model. 
Corresponding results for the $Z(2)$ model are given in the appendix.
The dashed curves are obtained from a truncated series expansion, given in Eq.\eqref{Ransatz}.
}
\label{fig:FSSzne0}
\end{figure}
Results for $H\chi_M/M$ versus $z$ for several values of $z_L$ are 
shown in Fig.~\ref{fig:FSSzne0}. From
these results, we may extract 
the intercepts $z_\delta(z_L)$.
As seen in the inset of Fig.~\ref{fig:FSSzne0},
these agree well with 
Eq.~\ref{zdelta2} 
for $|z|\le 0.6$ where 
$\widetilde{z}_L^2\,\lsim \,0.06$. 
We obtain for the FSS dependence of the critical temperature $T_c$,
\begin{eqnarray}
    T_c&=&T_\delta - \frac{z_\delta(z_L)}{z_0} H^{1/\beta\delta} 
    \nonumber\\
    &=&T_\delta +2 a_{G,2}\, \frac{ \beta \delta}{b_1}\,\frac{\widetilde{z}_{L,0}^2}{z_0}\, H^{1/\beta\delta-2}\, L^{2/\nu_c}  \; .
\label{Tc}
\end{eqnarray}
This suggest that the infinite volume limit at fixed $H$ is approached faster than $L^{-3}$, {\it i.e.} we find 
$T_\delta(H)-T_c \sim L^{2/\nu_c} = L^{4.96}$,
which clarifies the origin
of the large 
exponent of $L$ appearing in $z_\delta(z_L)$ that was observed previously also in the analysis of $T_\delta$ 
in calculations for the $O(4)$ model \cite{Engels:2014bra}.
It also provides a parametrically correct interpretation of the large
exponents found in the fit to
the finite-size dependence of
pseudo-critical temperatures extracted from the location of maxima in the thermal and magnetic susceptibilities in the $Z(2)$, $O(2)$ and $O(4)$
universality classes \cite{Karsch:2023pga}.

Having control of the parametric form of leading
and higher order finite-size 
correction to the infinite-volume scaling functions at non-vanishing symmetry breaking parameter $H$ also allows to eliminate 
a large contribution to systematic errors in the 
determination of the chiral phase transition temperature in $(2+1)$-flavor QCD \cite{HotQCD:2019xnw}. It also 
provides the basis for 
infinite volume extrapolation
at fixed $H$ as well as
a joint infinite-volume and vanishing $H$ extrapolation.
Eliminating the systematic
errors which in \cite{HotQCD:2019xnw} have been estimated on the basis
of $1/L^3$ extrapolations, 
reduces the quoted result of chiral transition
temperature within the quoted
errors by about $2$~MeV.

\section{Conclusions}
We determined that the finite-size corrections to the universal critical-isotherm ($J=J_c\Leftrightarrow z=0$) value $H\chi/M=1/\delta$ are proportional to ($1/(H L^{3/(1+1/\delta)}))^2$ in the leading order for both the $Z(2)$ and $O(2)$ spin models. We further showed that the two universality classes approach the asymptotic limit differently at smaller values of the finite-size scaling parameter $z_L<0.6$; while the leading order correction is proportional to $\widetilde{z}_L^2$ for O(2) model, both $\widetilde{z}_L^2$ and $\widetilde{z}_L^4$ terms are required for the $Z(2)$ model. We also find that the mixed term ($z\widetilde{z}_L^2$) is numerically negligible for $z_L<0.5$, considerably simplifying the finite-size scaling description in this regime. This finding is striking, as this fixes the leading finite-size correction entering in the determination of the chiral transition temperature of (2+1)-flavor QCD and thereby reduces the systematic uncertainty associated with the choice of an otherwise unknown finite-size ansatz. We also provide the leading and next to leading order parametric form of $f_G(0,z_L)$ for both the spin models.

\hspace{0.2cm}
\section*{Acknowledgments}
This work was supported by the Deutsche Forschungsgemeinschaft
(DFG, German Research Foundation) Proj. No. 315477589-TRR 211. Computations and data analysis required for this work have been performed on the GPU cluster at Bielefeld University, Germany, and
clusters Noctua2 and Otus at the NHR Center Paderborn Center for
Parallel Computing (PC2).
 We sincerely acknowledge 
 useful discussions with  Christian Schmidt.

\bibliography{bibFSS}

\appendix

\section{Finite-size fits and determination of \texorpdfstring{$H_0$}{H0} and \texorpdfstring{$L_0$}{L0}}\label{app:fit_details}
\subsection{Fit variables and general procedures}
We discuss here in a bit more detail, the fits performed to
determine the non-universal parameters entering the specific $Z(2)$ and $O(2)$ models studied by us. These 
parameters enter the definition of the scaling variables $z$ and $z_L$, which we split in a part depending only on the input
parameters of our numerical calculations - the bare scaling variables -- and a part depending on the non-universal scales. E.g.
the finite-size scaling variable is written as
\begin{equation}
    z_L = z_{L,0} \,z_{L,b}\; ,
    \label{eq:zL}
\end{equation}
with the bare finite-size scaling variable, $z_{L,b}$
and non-universal scale,
$z_{L,0}$ given by
\begin{equation}
  z_{L,b}=\frac{1}{L H^{\nu_c}}\;\;, \;\;
      z_{L,0} = L_0 H_0^{\nu_c}\; .
\end{equation}
All fits are performed in terms of $z_{L,b}$ prior to
the final determination of the scale factor.
The normalized variable $z_L$ used in the main text is introduced only after
the non-universal parameters $H_0$ and $L_0$ have been determined.

At $z=0$, we first determine the finite-size dependence of the ratio $H\chi_M/M$ from
fits over a specified range in $z_{L,b}$. 
As a fit ansatz we use
    \begin{equation}
  \frac{H\chi_M(0,z_{L,b})}{M(0,z_{L,b})}
  =
  \frac{1}{\delta}
  +\sum_{k=1}^{3} r_{2k,b}\,z_{L,b}^{\,k c_L},
  \label{eq:ratioO2}
\end{equation}
These fits determine the finite-size
dependence of the ratio of FSS functions of the order parameter ($f_G$) and its susceptibility ($f_\chi$), {\it i.e.} in the scaling region we have $H \chi/M=f_\chi(z,z_L)/f_G(z,z_L)$
without requiring any non-universal multiplicative 
factors. 

\begin{figure*}[htbp]
\includegraphics[width=0.45\textwidth]{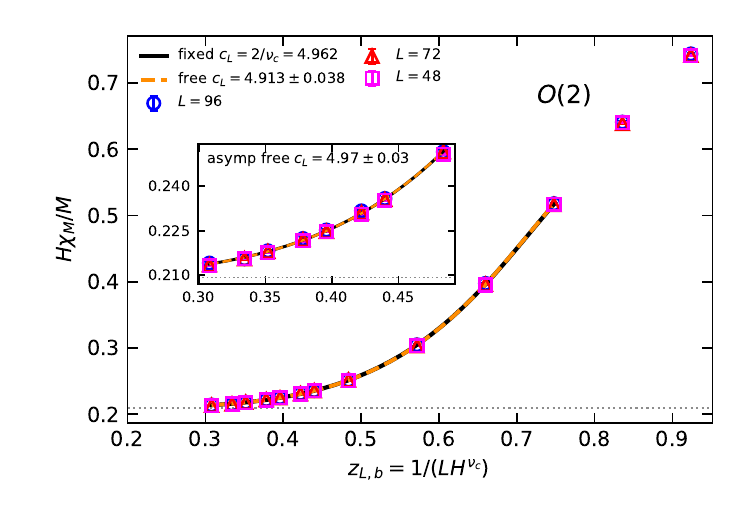}
\includegraphics[width=0.45\textwidth]{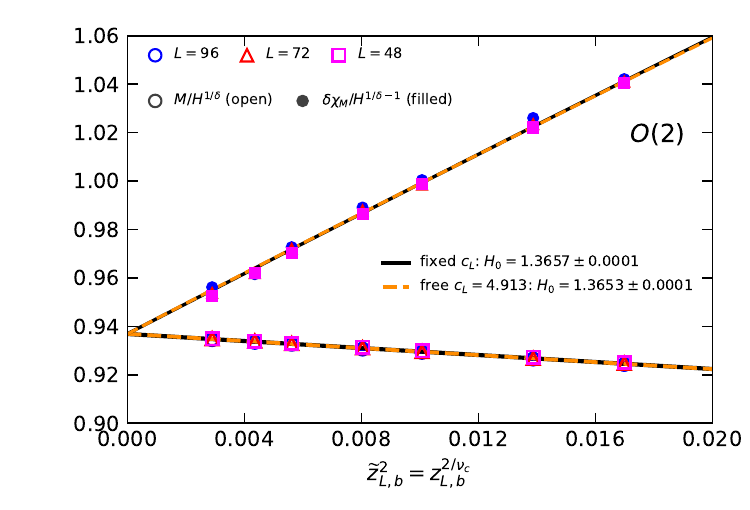}
\includegraphics[width=0.45\textwidth]{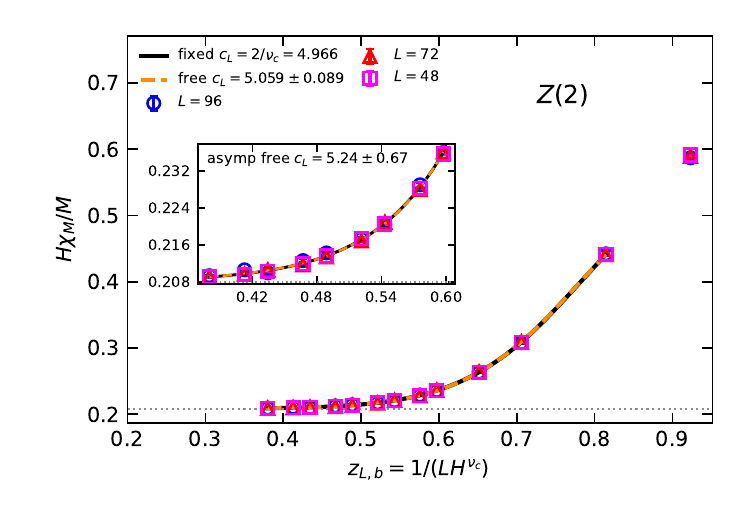}
\includegraphics[width=0.45\textwidth]{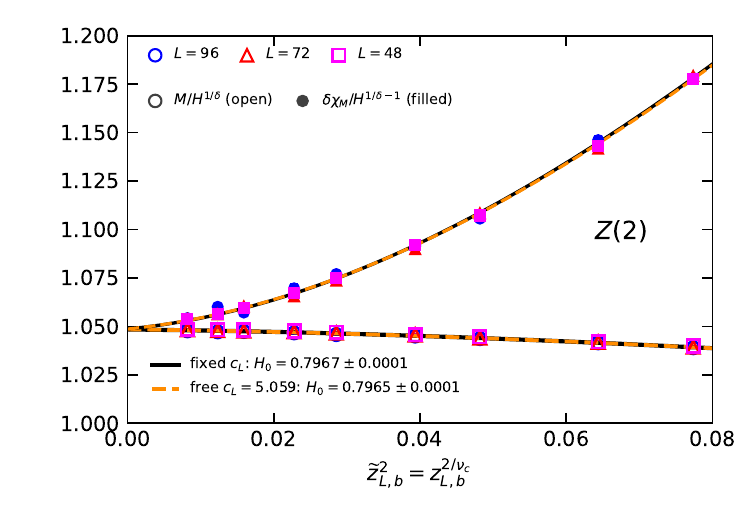}
\caption{({\it Left}:) The ratio $H\chi_M/M$ versus 
the bare scaling variable $z_{L,b}$ for the $3$-$d$, $O(2)$ (top) and $Z(2)$ (bottom) models.
({\it Right}:) The scaled order
parameter $M/H^{1/\delta}$ (open symbols) and the scaled susceptibility $\delta \chi_M/H^{1/\delta-1}$ (filled symbols)
versus 
the bare scaling variable $\widetilde{z}_{L,b}^2$ for the $3$-$d$, $O(2)$ (top) and $Z(2)$ (bottom) models.
}
\label{fig:Hchi_M_H0}
\end{figure*}

Having determined the expansion coefficients $r_{2k,b}$,
we can extract
from them the
series expansion for the FSS function of the order parameter,
\begin{equation}
    f_G(0,z_L) = 1 
    +\sum_{k=1}^{3} a_{G,2k}\,z_{L}^{\,k c_L} \label{eq:series-exp}
    \; .
\end{equation}
Using this ansatz and Eq.~\ref{fchiasym}
to determine the series 
for the FSS function $f_\chi(0,z_L)$, we can easily relate the expansion 
coefficients $r_{2k,b}$ and $a_{G.b}$,
\begin{eqnarray}
a_{G,2,b}&=&-r_{2,b}/2 \; , \\ 
a_{G,4,b}&=&r_{2,b}^2/8-r_{4,b}/4\; .
\end{eqnarray}
These fits are 
then used to extract the
normalization $H_0$ from
normalized order parameter,
$M/H^{1/\delta}$ by demanding for fixed $H$,
\begin{equation}
    \lim_{L\rightarrow\infty}\frac{M}{H^{1/\delta}} =
    H_0^{-1/\delta}\; .
\end{equation}
This fixes the normalization
condition for the order parameter scaling function
$f_G(0,0)=1$.
Having
determined $H_0$, we obtain $L_0$ from the normalization condition,
\begin{equation}
 \frac{f_G(0,z_L=1)}{f_G(0,0)}=0.7\; ,
 \label{L0norm}
\end{equation}
which defines $z_L=1$ as that value at which, the finite-size dependent value of the order parameter drops $30\%$ below its infinite volume limit value \cite{Karsch:2023pga}. 
Equivalently at fixed $H$,
the normalization condition Eq.~\ref{L0norm} defines
the volume $L^*$ or correspondingly the value $z_{L,b}^*$ at which 
\begin{equation}
    M(T_c,H,L^*)/H^{1/\delta} =0.7 H_0^{-1/\delta}\; . 
\end{equation}
Finally, the fitted coefficients appearing in the series expansion of $H\chi_M/M$ and $M$ in terms of $z_{L,b}^{c_L}$
are converted to coefficients in a series using the normalized variable $z_L^{c_L}$, 
    \begin{eqnarray}
  &&\frac{H\chi_M(0,z_{L,b})}{M(0,z_{L,b})}
  =
  \frac{1}{\delta}
  +\sum_{k=1}^{3} r_{2k}\,z_{L}^{\,k c_L}\; ,
  \label{eq:rationorm}
  \\
  &&M(0,z_{L,b}) = \left(\frac{H}{H_0} \right)^{1/\delta}\left(
  1+ \sum_{k=1}^{3} a_{G,2k}\,z_{L}^{\,k c_L}\right) \; ,
  \label{eq:Mnorm}
\end{eqnarray}
with
\begin{equation}
  r_{2k}
  =
  \frac{r_{2k,b}}{z_{L,0}^{\,k c_L}} \;\; ,\;\;
  a_{G,2k}=\frac{a_{G,2k,b}}{z_{L,0}^{\,k c_L}}
  \; .
  \label{r2knorm}
  \end{equation}
The series in terms of $z_L^{c_L}$ in Eq.~\ref{eq:eqntext} are used in the main text. 
Our data obtained in terms of
bare scaling variables are
shown in Fig.~\ref{fig:Hchi_M_H0}. 
Unless
stated otherwise, the fits use the $L=72$ and $96$ data sets, while the
$L=48$ results are shown for comparison. The results for the scale factors $H_0$ and $L_0$ as well as the
expansion coefficients $r_{2k}$ and $a_{G,2k}$ obtained from our fits, are
summarized in Table~\ref{tab:z2-o2-fits}.
\begin{table*}[t]
\centering
  \renewcommand{\arraystretch}{1.12}
  \begin{tabular*}{0.85\textwidth}{@{\extracolsep{\fill}}c  c  c}
    \hline\hline
Scale factors and expansion coefficients& O(2) & Z(2) \\
    \hline \hline
    $H_0$ & $1.36574(15)(63)$ & $0.79665(11)(88)$ \\ \hline
    $L_0$ & $0.9849(3)(21)$ & $1.0239(9)(7)$ \\ \hline
    $r_2$ & $0.9019(39)(94)$ & $0.1086(48)(4)$ \\ \hline
    $r_4$ & $-0.087(36)(2)$ & $8.66(14)(6)$ \\
    \hline\hline\hline
    $z_{L,0}=L_0H_0^{\nu_c}$ & $1.1167(3)(24)$ & $0.9343(8)(6)$ \\ \hline
    $a_{G,2}=-r_2/2$
      & $-0.4509(20)(47)$ & $-0.0543(24)(144)$ \\ \hline
    $a_{G,4}=r_2^2/8-r_4/4$ & $0.123(9)$ & $-2.16(3)$ 
    \\
    \hline\hline
  \end{tabular*}
    \caption{Fit results for the scale factors $H_0$ and $L_0$ as well as the
    leading and next-to-leading order expansion coefficients $r_{2k}$ 
    for the $3$-$d$,$O(2)$
    and $Z(2)$ universality classes.
  Parentheses on fitted quantities give statistical and, where
  present, systematic errors in that order. The lower part lists the scale parameters and expansion coefficients derived from those listed in the upper part, of the table.}
  \label{tab:z2-o2-fits}
\end{table*}
Further details on the fits
performed for the determination of the scale factors $H_0$ and $L_0$ as 
well as the expansion coefficients $r_{2k}$ are
given in the following subsections 
separately for the $3$-$d$, $O(2)$ and $Z(2)$
models.

\subsection{\texorpdfstring{$\mathbf{O(2)}$}{O2} case}

\subsubsection{Determination of \texorpdfstring{$r_{2,b}$}{r2b}}
At $z=0$, we fit $H\chi_M/M$ using the ansatz given in Eq.(\ref{eq:ratioO2}), in the range
$z_{L,b}\in[0.30,0.75]$. Fixing
$\widetilde c_L=\nu_c c_L=2$ gives
$\chi^2/\mathrm{dof}=3.09$, while a free fit gives
$\widetilde c_L=1.980(15)$ with
$\chi^2/\mathrm{dof}=3.16$, consistent with the expected value. Restricting the fit to $z_{L,b}\in[0.30,0.50]$, retaining only the leading  term ($k=1$) in
Eq.~\eqref{eq:ratioO2} for fixed value gives $\chi^2/\mathrm{dof}=2.27$, while a free fit gives
$\widetilde c_L=2.005(12)$ with $\chi^2/\mathrm{dof}=2.43$. The fitted values of the
exponent are therefore consistent with the expected value in both fit
ranges.
The leading
coefficient in the bare variable is
\begin{equation}
  a_{G,2,b}
  =
  -\frac{r_{2,b}}{2}
  =
  -0.7798(32).
\end{equation}
The fit over $z_{L,b}\in[0.30,0.50]$ gives
$a_{G,2,b}=-0.781(3)$, consistent with the result from the wider fit range.

\subsubsection{Determination of \texorpdfstring{$H_0$}{H0} }
With the finite-size dependence of $f_G$ fixed, a fit to the magnetization over $z_{L,b}\in[0.30,0.50]$ gives
\begin{equation}
  H_0=1.36574(15).
\end{equation}
A corresponding fit to
$\delta\chi_M/H^{1/\delta-1}$ gives $H_0=1.3651(16)$. Taking their difference as a systematic uncertainty, we obtain
\begin{equation}
  H_0=1.36574(15)_{\rm stat}(63)_{\rm sys}.
\end{equation}
This agrees within $0.04\%$ with $H_0=1.36632$ quoted in \cite{Karsch:2023pga} . Variations of
the fit range and of the truncation used in fits to $H\chi_M/M$ give changes smaller
than this systematic uncertainty.

\subsubsection{Determination of \texorpdfstring{$L_0$}{L0}}
We determine $L_0$ using the normalization condition in Eq.~\eqref{L0norm}.

\begin{figure*}[htbp]
  \centering
  \includegraphics[width=0.45\textwidth]{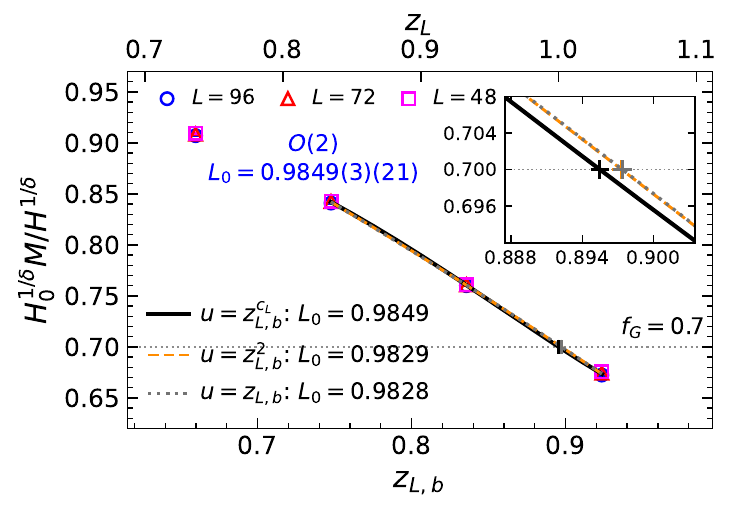}
  \includegraphics[width=0.45\textwidth]{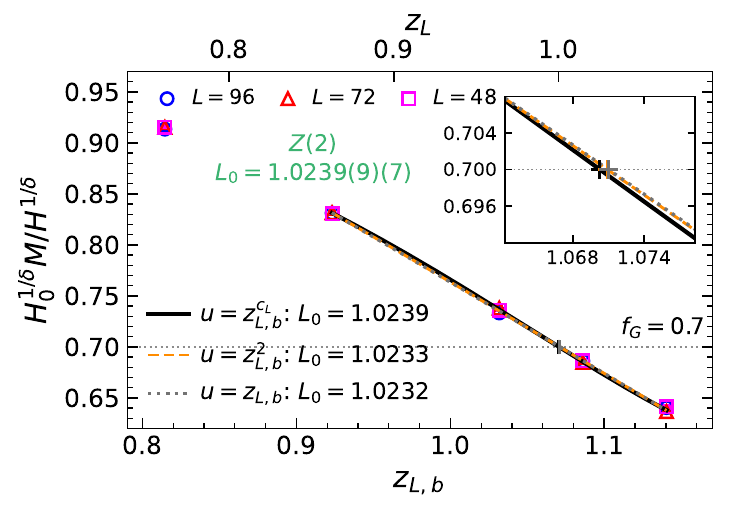}
  \caption{Results for
  $f_G=H_0^{1/\delta}M/H^{1/\delta}$ at $z=0$ near the
    normalization point $f_G(0,z_L)=0.7$, for $L=96$, $72$, $48$ (open symbols) obtained in 
    calculations for the $3$-$d$,
  $O(2)$ model ({\it left}) and $Z(2)$ model ({\it right}), respectively.
    Lines show fits with the ansatz $f_G=c_0+c_1u+c_2u^2$ fitted to data on $L=72$, $96$ lattices in the interval
    $z_L\in[0.85,1.10]$
    in the case of $O(2)$ and
    $z_{L,b}\in[0.923,1.140]$
    in the case of $Z(2)$.
    For the fit variable $u$,
    three choices have been 
    used:
    $u=\widetilde{z}_L^{\,2}$ (solid, quoted
    value), $u=z_{L,b}^2$ (dashed) and $u=z_{L,b}$ (dotted). Also shown is 
    $H\chi_M/M$ obtained from
    the twelfth-order fit to
    $f_G(0,z_L)$ 
    (dash-dotted). The inset zooms on the crossings with $f_G=0.7$; each line's $L_0$ is given in
    the legend.
The resulting fit values for $L_0$ are given
in the legend.
}
\label{fig:L0local}
\end{figure*}

As shown in Fig.~\ref{fig:L0local} (left), we fit the data over  $z_{L,b}\in[0.75,0.94]$ using
\begin{equation}
  f_G=c_0+c_1u+c_2u^2,
  \qquad
  u=z_{L,b}^{\,c_L}.
  \label{eq:L0fitO2}
\end{equation}
This gives
\begin{equation}
  z_{L,b}^{*}=0.8955,
\end{equation}
and hence
\begin{align}
  z_{L,0}
  &=1.1167(3)_{\rm stat}(24)_{\rm sys},
  \\
  L_0
  &=0.9849(3)_{\rm stat}(21)_{\rm sys}.
  \label{eq:L0resultO2}
\end{align}
The statistical uncertainty includes the uncertainty in $H_0$.

To estimate the systematic uncertainty, we repeat the quadratic interpolation
using $u=z_{L,b}^{2}$ and $u=z_{L,b}$, which give $L_0=0.9829$ and
$0.9828$, respectively. The largest deviation from the central value is taken
as the systematic uncertainty. Straight-line fits in each of the three
variables do not describe the three data points in this range, with
$\chi^2/\mathrm{dof}$ between approximately $10$ and $440$. Our result is
about $0.6\%$ larger than the value $L_0=0.97917$ obtained in \cite{Karsch:2023pga} 
from a fit to $f_G(0,z_{L,b})$
in a large fit range.

\subsubsection{Normalized coefficients.}
Once $z_{L,0}$ is known, we obtain the expansion coefficients in the normalized variable 
from Eq.~\ref{r2knorm}
with $z_{L,0}^{\,c_L}=1.7294$  as,
\begin{align}
  a_{G,2}=-0.4509(20) \; ,\ a_{G,4}=0.123(9)\; ,\ a_{G,6}=0.09(2) \; .
\end{align}

\subsection{\texorpdfstring{$\mathbf{Z(2)}$}{Z2} case}

For the $Z(2)$ analysis, we use the critical exponents quoted also in \cite{Karsch:2023pga},
\begin{equation}
  \delta=4.805,
  \qquad
  \beta=0.3258,
  \qquad
\nu_c =(1+1/\delta)/3\; .
\end{equation}
The determination of the non-universal scale factors $H_0$ and $L_0$ proceeds in 
complete analogy to the one
discussed for the $O(2)$ case.

\subsubsection{Finite-size scaling of \texorpdfstring{$H\chi_M/M$}{HchiMM}}
At $z=0$, we use the same finite-size expansion as in the $O(2)$ analysis. Over $z_{L,b}\in[0.38,0.92]$, a truncation at $n=3$ fails to describe the
data, giving $\chi^2/\mathrm{dof}=40.5$ 
for fixed $c_L=1/\nu_c$.
Including the next term,
$n=4$, reduces this to $\chi^2/\mathrm{dof}=1.98$. Allowing the exponent to vary gives
\begin{equation}
  \widetilde c_L=2.037(36),
\end{equation}
with $\chi^2/\mathrm{dof}=2.01$, consistent with the expected value.

In the smaller range, $z_{L,b}\in[0.38,0.60]$, the first two
terms in Eq.~\eqref{eq:ratioO2} are sufficient. The fixed-exponent fit gives
$\chi^2/\mathrm{dof}=2.08$, while a free fit gives $\widetilde c_L=2.11(27)$. We obtain
$a_{G,2,b}=-0.0490(22)$ and use the difference between the two
determinations as the systematic uncertainty on the leading coefficient.

\subsubsection{Determination of \texorpdfstring{$H_0$}{H0}}
With the finite-size dependence of $f_G$ fixed by the $n=4$ fit, $H_0$ is
determined from a fit in the range $z_{L,b}\in[0.380,0.597]$. This gives
\begin{equation}
  H_0=0.79665(11).
\end{equation}
As a consistency check, the corresponding fit to
$\delta\chi_M/H^{1/\delta-1}$ gives $H_0=0.7958(9)$. We use the
magnetization result as the central value and take the difference between the
two determinations as a systematic uncertainty,
\begin{equation}
  H_0=0.79665(11)_{\rm stat}(88)_{\rm sys}.
\end{equation}
This value is about $0.18\%$ larger than the value $H_0=0.79522$ obtained in \cite{Karsch:2023pga}.

\subsubsection{Determination of \texorpdfstring{$L_0$}{L0}}
As in the $O(2)$ case, we determine $L_0$ from the normalization
condition~\cite{Karsch:2023pga}. Data close to $f_G(0,z_L=1)$
are shown in Fig.~\ref{fig:L0local}~(right).

The crossing lies between the data at $z_{L,b}=1.032$ and $1.086$. We
therefore determine it from the magnetization data over
$z_{L,b}\in[0.923,1.140]$, which covers the region where $f_G$ approaches
the normalization condition $f_G=0.7$. We fit
\begin{equation}
  f_G=c_0+c_1u+c_2u^2,
  \qquad
  u=z_{L,b}^{\,c_L}.
  \label{eq:L0fitZ2}
\end{equation}
This gives
\begin{equation}
  z_{L,b}^{*}=1.0704,
\end{equation}
and hence
\begin{align}
  z_{L,0}
  &=0.9343(8)_{\rm stat}(6)_{\rm sys},
  \\
  L_0
  &=1.0238(9)_{\rm stat}(7)_{\rm sys}.
  \label{eq:L0resultZ2}
\end{align}

\begin{table*}[t]
\label{tab:z2-o2-comparison}
  \centering
  \begingroup
  \renewcommand{\arraystretch}{1.12}
  \begin{tabular*}{0.85\textwidth}{@{\extracolsep{\fill}}ccc}
    \hline\hline
    critical exponent and fit details & $O(2)$ & $Z(2)$ \\
    \hline\hline
    $\delta$ & $4.7798$ & $4.805$ \\ \hline
    $\beta$ & $0.34864$ & $0.3258$ \\ \hline
    $\nu_c$ & $0.40307$ & $0.40271$ \\ \hline
    Fixed $c_L=2/\nu_c$ & $4.96190$ & $4.96641$ \\ \hline
    $H\chi_M/M$ fit window in $z_L$ & $[0.35,0.85]$ & $[0.35,0.85]$ \\ \hline
    Fits with order $2n$ polynomial &  $n=3$ & $n=4$ \\ \hline
    $\chi^2/\mathrm{dof}$, fixed $c_L$ & $3.09$ & $1.98$ \\ \hline
    $c_L$ from ''free fit'' & $4.913(38)$ & $5.059(89)$ \\ \hline
    Small-$z_L$ window & $[0.35,0.55]$ & $[0.35,0.55]$ \\ \hline
    Small-$z_L$ terms & $1$ & $2$ \\ \hline
    Small-$z_L$ $\chi^2/\mathrm{dof}$, fixed $c_L$ & $2.27$ & $2.08$ \\
    \hline\hline\hline
    $f_G(z,z_L=0)=1+b_1z+b_2z^2$&~&~ \\ \hline \hline \hline
    $b_1$ & $-0.259(2)$ & $-0.21866(1)$ \\ \hline
    $b_2$ & $-0.026(1)$ & $-0.0177767(5)$ \\
    \hline\hline
  \end{tabular*}
    \caption{Comparison of the $O(2)$ and $Z(2)$ analyses at $z=0$ and the
  small-$z$ expansion. Fits use $L=72,96$.
  Here $z_{L,b}=1/(N_s H^{\nu_c})$, $z_L=z_{L,0}z_{L,b}$,
  $\widetilde z_L=z_L^{1/\nu_c}$, and $c_L=2/\nu_c$.
  Parentheses on fitted quantities give statistical errors.}
  \endgroup
\end{table*}

To estimate the systematic uncertainty, we repeat the quadratic interpolation
using $u=z_{L,b}^{2}$ and $u=z_{L,b}$, which give
$L_0=1.0232$ and $1.0231$, respectively. The largest deviation from the
central value is taken as the systematic uncertainty. Our result is about
$0.23\%$ smaller than the value $L_0=1.0262$ obtained in \cite{Karsch:2023pga}.
\begin{figure}[t]
\includegraphics[width=0.45\textwidth]{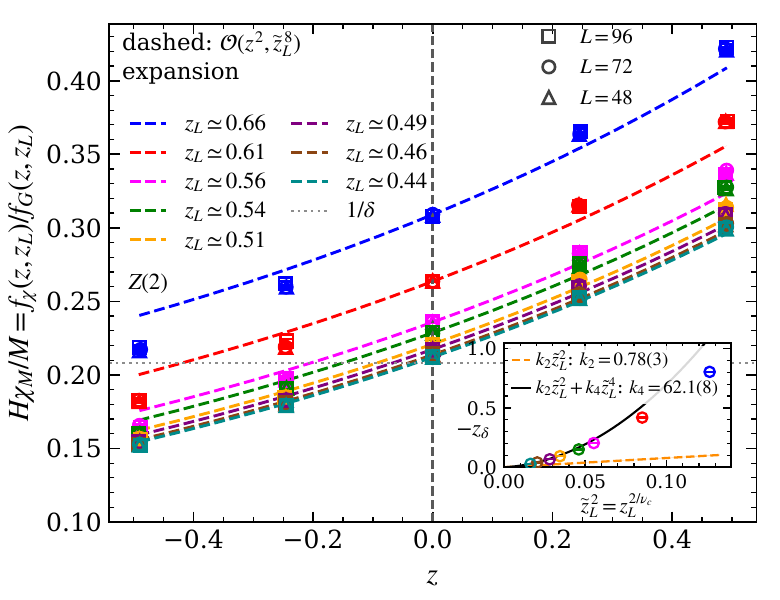}
\caption{The ratio $H\chi_M/M=\frac{f_\chi(z,z_L)}{f_G(z,z_L)}$ versus $z$ for selected $z_L$ values.
Shown are results for the $3$-$d$, $Z(2)$ model. The dashed curves obtained from a truncated series expansion, given in Eq.(\ref{Ransatz}).
}
\label{figZ2:FSSzne0}
\end{figure}

\subsubsection{Normalized coefficients.}
With $z_{L,0}$ known, the expansion coefficients are converted to the normalized
finite-size variable gives
\begin{eqnarray}
  a_{G,2}
  &=&
  -0.0543(24) \; ,\ a_{G,4}=-2.17(3)\; , \notag \\ 
  a_{G,6}
  &=&
  4.98(13)\; ,\ a_{G,8}=-1.60(7)\; .
\end{eqnarray}

\section{Numerical FSS analysis for \texorpdfstring{$z\neq0$}{z0} for the \texorpdfstring{$\mathbf{Z(2)}$}{Z2} case}
Results for $H\chi_M/M$ versus $z$ in the case for $O(2)$ universality class are already
shown in Fig.~\ref{fig:FSSzne0}, for several values of $z_L$. Here, we do the same for $Z(2)$ class, and as seen in the inset of Fig.~\ref{figZ2:FSSzne0},
these agree well with 
Eq.~\ref{zdelta2} 
for $|z|\le 0.6$ where 
$\widetilde{z}_L^2\,\lsim \,0.06$. From
these results, we may extract 
the intercepts $z_\delta(z_L)$.

\section{Simulation details}
For each parameter set, we generated $N_{\rm st}$ independent Markov-chain
streams and performed measurements after each complete update cycle.
One update cycle consists of $25$ cluster updates followed by $100$
Metropolis sweeps for the $Z(2)$ model, and $25$ cluster updates followed
by $25$ Metropolis sweeps for the $O(2)$ model. The ensembles listed in
Tables~\ref{tab:o2-simulation-statistics} and 
\ref{tab:z2-simulation-statistics} are generated at the critical temperature $J_c$, and hence correspond to $t=0$ or equivalently $z=0$. Similar ensembles have also been generated at non-critical couplings, $J \neq J_c$, where coupling parameters $J$ and $H$ are varied ensuring fixed $z$ and $z_{L}$. These ensembles are tabulated in Tables~\ref{tab:o2-simulation-statistics-nonzero} and \ref{tab:z2-simulation-statistics-nonzero}.  


\begin{table*}[t]
\caption{
Simulation parameters and statistics for the $O(2)$ model at the critical
coupling $J_c=0.5091463(10)$, corresponding to $z=0$.
$N_{\rm str}$ is the number of independent Markov-chain streams and
$N_{\rm meas}$ is the total number of magnetization measurements, summed over
all streams, that enter the analysis after thermalization.
}
\label{tab:o2-simulation-statistics}

\begin{ruledtabular}
\begin{tabular}{
ccc @{\hspace{10pt}}|| 
ccc @{\hspace{10pt}}|| 
ccc
}
\multicolumn{3}{c||}{$L=48$} &
\multicolumn{3}{c||}{$L=72$} &
\multicolumn{3}{c}{$L=96$}
\\ \hline \hline
$H$ & $N_{\rm str}$ & $N_{\rm meas}$ &
$H$ & $N_{\rm str}$ & $N_{\rm meas}$ &
$H$ & $N_{\rm str}$ & $N_{\rm meas}$
\\
\hline\hline

0.00183749  & 10 & 15000000 &
0.000672020 & 10 & 8592900  &
0.000329183 & 10 & 4440000  \\

0.00125356  & 10 & 15000000 &
0.000458461 & 10 & 8632950  &
0.000224573 & 10 & 4563000  \\

0.00102222  & 10 & 15000000 &
0.000373852 & 10 & 8662950  &
0.000183128 & 10 & 4695000  \\

0.000900078 & 10 & 15000000 &
0.000329183 & 10 & 8644950  &
0.000161247 & 10 & 4743000  \\

0.000752251 & 10 & 15000000 &
0.000275119 & 10 & 8688900  &
0.000134764 & 10 & 4779900  \\

0.000672020 & 10 & 15000000 &
0.000245776 & 10 & 8727900  &
0.000120391 & 10 & 4990950  \\

0.000572597 & 10 & 15000000 &
0.000209414 & 10 & 8745000  &
0.000102580 & 10 & 4951950  \\ \hline

0.000517450 & 10 & 15000000 &
0.000189245 & 10 & 8736000  &
0.0000927002 & 10 & 4603950 \\

0.000408491 & 10 & 15000000 &
0.000149396 & 10 & 8853000  &
0.0000731804 & 10 & 4647900 \\

0.000269898 & 10 & 15000000 &
0.0000987092 & 10 & 8868900 &
0.0000483518 & 10 & 4641900 \\

0.000189245 & 10 & 15000000 &
0.0000692122 & 10 & 8916000 &
0.0000339030 & 10 & 4647900 \\

0.000138732 & 10 & 15000000 &
0.0000507381 & 10 & 9163950 &
0.0000248536 & 10 & 4758900 \\

0.000105279 & 10 & 15000000 &
0.0000385036 & 10 & 9168000 &
0.0000188606 & 10 & 5076000 \\

0.0000821325 & 10 & 15000000 &
0.0000300381 & 10 & 8829000 &
0.0000147139 & 10 & 5073000 \\

\end{tabular}
\end{ruledtabular}
\end{table*}

\begin{table*}[t]
\caption{
Simulation parameters and statistics for the $Z(2)$ model at the critical
coupling $J=J_c=0.3750966(4)$, corresponding to $z=0$.
$N_{\rm str}$ is the number of independent Markov-chain streams and
$N_{\rm meas}$ is the total number of magnetization measurements, summed over
all streams.
}
\label{tab:z2-simulation-statistics}

\begin{ruledtabular}
\begin{tabular}{
ccc @{\hspace{10pt}}|| 
ccc @{\hspace{10pt}}|| 
ccc
}
\multicolumn{3}{c||}{$L=48$} &
\multicolumn{3}{c||}{$L=72$} &
\multicolumn{3}{c}{$L=96$}
\\ \hline \hline
$H$ & $N_{\rm str}$ & $N_{\rm meas}$ &
$H$ & $N_{\rm str}$ & $N_{\rm meas}$ &
$H$ & $N_{\rm str}$ & $N_{\rm meas}$
\\
\hline \hline

0.00108292  & 4  & 10259580 &
0.000395648 & 4  & 7515180  &
0.000193664 & 4  & 5277960  \\

0.000738495 & 10 & 27655600 &
0.000269811 & 10 & 13187800 &
0.000132069 & 10 & 7016416  \\

0.000602081 & 10 & 25656400 &
0.000219972 & 10 & 11489800 &
0.000107673 & 10 & 2725968  \\

0.000530073 & 10 & 27655600 &
0.000193664 & 10 & 13244400 &
0.0000947956 & 10 & 6956617 \\

0.000442934 & 10 & 25656400 &
0.000161827 & 10 & 11433200 &
0.0000792121 & 10 & 2386248 \\

0.000395648 & 10 & 27655600 &
0.000144551 & 10 & 13244400 &
0.0000707557 & 10 & 7594473 \\

0.000337058 & 10 & 23011464 &
0.000123145 & 10 & 7768386 &
0.0000602778 & 10 & 4006533 \\

0.000304565 & 10 & 27655600 &
0.000111274 & 10 & 12565200 &
0.0000544668 & 10 & 6518091 \\

0.000263534 & 10 & 23103360 &
0.0000962831 & 10 & 7827840 &
0.0000471291 & 10 & 3388610 \\ \hline

0.000240375 & 10 & 26156200 &
0.0000878216 & 10 & 12565200 &
0.0000429874 & 10 & 6139364 \\

0.000193664 & 10 & 24990000 &
0.0000707557 & 10 & 12565200 &
0.0000346339 & 10 & 6099498 \\

0.000158754 & 10 & 20297130 &
0.0000580011 & 10 & 12565200 &
0.0000283907 & 10 & 2075088 \\

0.000111274 & 4 & 3199200 &
0.0000406541 & 4 & 795960 &
0.0000198996 & 4 & 318780 \\

0.0000815465 & 4 & 3199200 &
0.0000297933 & 4 & 795960 &
0.0000145834 & 4 & 317580 \\

0.0000618658 & 4 & 3199200 &
0.0000226028 & 4 & 756360 &
0.0000110638 & 4 & 317580 \\

0.0000544668 & 4 & 799200 &
0.0000198996 & 4 & 795960 &
0.00000974056 & 4 & 764760 \\

0.0000482517 & 4 & 799200 &
0.0000176289 & 4 & 795960 &
0.00000862908 & 4 & 647580 \\

0.0000384947 & 4 & 799200 &
0.0000140641 & 4 & 795960 &
0.00000688419 & 4 & 750780 \\

\end{tabular}
\end{ruledtabular}
\end{table*}

\begin{table*}[t]
\caption{
Simulation parameters and statistics for the $O(2)$ model away from the
critical coupling ($z\neq0$).
The four blocks in the Table 
correspond to $z=-0.52$,
$-0.26$, $0.26$, $0.52$.
$J$ and $H$ are 
the inverse temperature and external field couplings
of each
ensemble; within each volume, ensembles generated at the same value of $z$
are grouped between horizontal lines.
$N_{\rm str}$ is the number of independent Markov-chain streams and
$N_{\rm meas}$ is the total number of magnetization measurements, summed over
all streams, that enter the analysis after thermalization.
}
\label{tab:o2-simulation-statistics-nonzero}

\begin{ruledtabular}
\begin{tabular}{
cccc @{\hspace{10pt}} ||
cccc @{\hspace{10pt}} ||
cccc
}

\multicolumn{4}{c||}{$L=48$} &
\multicolumn{4}{c||}{$L=72$} &
\multicolumn{4}{c}{$L=96$}
\\ \hline \hline

$J$ & $H$ & $N_{\rm str}$ & $N_{\rm meas}$ &
$J$ & $H$ & $N_{\rm str}$ & $N_{\rm meas}$ &
$J$ & $H$ & $N_{\rm str}$ & $N_{\rm meas}$
\\
\hline\hline

0.510412 & 0.000672020 & 10 & 999900 &
0.509840 & 0.000245776 & 10 & 1999950 &
0.509600 & 0.000120391 & 10 & 1999950 \\

0.510296 & 0.000572597 & 10 & 999900 &
0.509777 & 0.000209414 & 10 & 1999950 &
0.509559 & 0.000102580 & 10 & 1999950 \\

0.510139 & 0.000447806 & 10 & 999900 &
0.509691 & 0.000163775 & 10 & 1999950 &
0.509503 & 0.0000802237 & 10 & 1999950 \\

0.510086 & 0.000408491 & 10 & 999900 &
0.509662 & 0.000149396 & 10 & 1999950 &
0.509484 & 0.0000731804 & 10 & 1999950 \\

0.509880 & 0.000269898 & 10 & 999900 &
0.509549 & 0.0000987092 & 10 & 1999950 &
0.509411 & 0.0000483518 & 10 & 1999950 \\

\hline

0.509780 & 0.000672020 & 10 & 999900 &
0.509495 & 0.000245776 & 10 & 1999950 &
0.509375 & 0.000120391 & 10 & 1999950 \\

0.509723 & 0.000572597 & 10 & 999900 &
0.509464 & 0.000209414 & 10 & 1999950 &
0.509355 & 0.000102580 & 10 & 1999950 \\

0.509644 & 0.000447806 & 10 & 999900 &
0.509421 & 0.000163775 & 10 & 1999950 &
0.509327 & 0.0000802237 & 10 & 1999950 \\

0.509618 & 0.000408491 & 10 & 999900 &
0.509406 & 0.000149396 & 10 & 1999950 &
0.509317 & 0.0000731804 & 10 & 1999950 \\

0.509515 & 0.000269898 & 10 & 999900 &
0.509350 & 0.0000987092 & 10 & 1999950 &
0.509281 & 0.0000483518 & 10 & 1999950 \\

\hline

0.508523 & 0.000672020 & 10 & 999900 &
0.508807 & 0.000245776 & 10 & 1999950 &
0.508926 & 0.000120391 & 10 & 1999950 \\

0.508580 & 0.000572597 & 10 & 999900 &
0.508838 & 0.000209414 & 10 & 1999950 &
0.508947 & 0.000102580 & 10 & 1999950 \\

0.508658 & 0.000447806 & 10 & 999900 &
0.508881 & 0.000163775 & 10 & 1999950 &
0.508975 & 0.0000802237 & 10 & 1999950 \\

0.508684 & 0.000408491 & 10 & 999900 &
0.508895 & 0.000149396 & 10 & 1999950 &
0.508984 & 0.0000731804 & 10 & 1999950 \\

0.508787 & 0.000269898 & 10 & 999900 &
0.508952 & 0.0000987092 & 10 & 1999950 &
0.509021 & 0.0000483518 & 10 & 1999950 \\

\hline

0.507896 & 0.000672020 & 10 & 999900 &
0.508463 & 0.000245776 & 10 & 1999950 &
0.508702 & 0.000120391 & 10 & 1999950 \\

0.508010 & 0.000572597 & 10 & 999900 &
0.508526 & 0.000209414 & 10 & 1999950 &
0.508743 & 0.000102580 & 10 & 1999950 \\

0.508166 & 0.000447806 & 10 & 999900 &
0.508612 & 0.000163775 & 10 & 1999950 &
0.508799 & 0.0000802237 & 10 & 1999950 \\

0.508219 & 0.000408491 & 10 & 999900 &
0.508641 & 0.000149396 & 10 & 1999950 &
0.508818 & 0.0000731804 & 10 & 1999950 \\

0.508424 & 0.000269898 & 10 & 999900 &
0.508753 & 0.0000987092 & 10 & 1999950 &
0.508891 & 0.0000483518 & 10 & 1999950 \\
\end{tabular}
\end{ruledtabular}
\end{table*}

\begin{table*}[t]
\caption{
Simulation parameters and statistics for the $Z(2)$ model away from the
critical coupling ($J\neq J_c$). 
The four blocks in the Table 
correspond to $z=-0.49$,
$-0.25$, $0.25$, $0.49$.
$J$ and $H$ are 
the inverse temperature and external field couplings,
respectively; $N_{\rm str}$ is the number of independent Markov-chain streams
and $N_{\rm meas}$ is the total number of magnetization measurements summed
over all streams.
}
\label{tab:z2-simulation-statistics-nonzero}

\begin{ruledtabular}
\begin{tabular}{
cccc @{\hspace{10pt}}||
cccc @{\hspace{10pt}}||
cccc
}

\multicolumn{4}{c||}{$L=48$} &
\multicolumn{4}{c||}{$L=72$} &
\multicolumn{4}{c}{$L=96$}
\\ \hline \hline
$J$ & $H$ & $N_{\rm str}$ & $N_{\rm meas}$ &
$J$ & $H$ & $N_{\rm str}$ & $N_{\rm meas}$ &
$J$ & $H$ & $N_{\rm str}$ & $N_{\rm meas}$
\\
\hline
\hline

0.375562 & 0.000442934 & 10 & 9990000 &
0.375341 & 0.000161827 & 10 & 3990000 &
0.375251 & 0.0000792121 & 10 & 2989950 \\

0.375529 & 0.000395648 & 4 & 4537560 &
0.375324 & 0.000144551 & 4 & 1540380 &
0.375241 & 0.0000707557 & 4 & 597600 \\

0.375487 & 0.000337058 & 10 & 9990000 &
0.375302 & 0.000123145 & 10 & 3990000 &
0.375227 & 0.0000602778 & 10 & 2989950 \\

0.375463 & 0.000304565 & 4 & 4582800 &
0.375289 & 0.000111274 & 4 & 1410360 &
0.375218 & 0.0000544668 & 4 & 621180 \\

0.375430 & 0.000263534 & 10 & 9990000 &
0.375272 & 0.0000962831 & 10 & 3990000 &
0.375208 & 0.0000471291 & 10 & 2989950 \\

0.375411 & 0.000240375 & 4 & 4614000 &
0.375262 & 0.0000878216 & 4 & 1425180 &
0.375201 & 0.0000429874 & 4 & 622380 \\

0.375371 & 0.000193664 & 4 & 4523160 &
0.375241 & 0.0000707557 & 4 & 1427580 &
0.375188 & 0.0000346339 & 4 & 621180 \\

0.375338 & 0.000158754 & 4 & 4534800 &
0.375223 & 0.0000580011 & 4 & 1430760 &
0.375177 & 0.0000283907 & 4 & 625200 \\

\hline

0.375329 & 0.000442934 & 10 & 9990000 &
0.375219 & 0.000161827 & 10 & 3990000 &
0.375174 & 0.0000792121 & 10 & 2989950 \\

0.375313 & 0.000395648 & 4 & 4419960 &
0.375210 & 0.000144551 & 4 & 1539960 &
0.375169 & 0.0000707557 & 4 & 599160 \\

0.375292 & 0.000337058 & 10 & 9990000 &
0.375199 & 0.000123145 & 10 & 3990000 &
0.375162 & 0.0000602778 & 10 & 2989950 \\

0.375280 & 0.000304565 & 4 & 4444380 &
0.375193 & 0.000111274 & 4 & 1419600 &
0.375158 & 0.0000544668 & 4 & 619200 \\

0.375263 & 0.000263534 & 10 & 9990000 &
0.375184 & 0.0000962831 & 10 & 3990000 &
0.375152 & 0.0000471291 & 10 & 2989950 \\

0.375254 & 0.000240375 & 4 & 4470000 &
0.375179 & 0.0000878216 & 4 & 1428780 &
0.375149 & 0.0000429874 & 4 & 624000 \\

0.375234 & 0.000193664 & 4 & 4489200 &
0.375169 & 0.0000707557 & 4 & 1430400 &
0.375142 & 0.0000346339 & 4 & 622380 \\

0.375217 & 0.000158754 & 4 & 4508760 &
0.375160 & 0.0000580011 & 4 & 1486760 &
0.375137 & 0.0000283907 & 4 & 625560 \\

\hline

0.374864 & 0.000442934 & 10 & 9990000 &
0.374974 & 0.000161827 & 10 & 3990000 &
0.375019 & 0.0000792121 & 10 & 2989950 \\

0.374881 & 0.000395648 & 4 & 4477560 &
0.374983 & 0.000144551 & 4 & 1504440 &
0.375025 & 0.0000707557 & 4 & 603960 \\

0.374902 & 0.000337058 & 10 & 9990000 &
0.374994 & 0.000123145 & 10 & 3990000 &
0.375032 & 0.0000602778 & 10 & 2989950 \\

0.374914 & 0.000304565 & 4 & 4799160 &
0.375000 & 0.000111274 & 4 & 1429980 &
0.375036 & 0.0000544668 & 4 & 622380 \\

0.374930 & 0.000263534 & 10 & 9990000 &
0.375009 & 0.0000962831 & 10 & 3990000 &
0.375041 & 0.0000471291 & 10 & 2989950 \\

0.374939 & 0.000240375 & 4 & 4799160 &
0.375014 & 0.0000878216 & 4 & 1439160 &
0.375044 & 0.0000429874 & 4 & 631560 \\

0.374960 & 0.000193664 & 4 & 4799160 &
0.375025 & 0.0000707557 & 4 & 1438800 &
0.375051 & 0.0000346339 & 4 & 625980 \\

0.374976 & 0.000158754 & 4 & 4799160 &
0.375033 & 0.0000580011 & 4 & 1447200 &
0.375056 & 0.0000283907 & 4 & 663138 \\

\hline

0.374632 & 0.000442934 & 10 & 9990000 &
0.374852 & 0.000161827 & 10 & 3990000 &
0.374942 & 0.0000792121 & 10 & 2989950 \\

0.374665 & 0.000395648 & 4 & 4549560 &
0.374869 & 0.000144551 & 4 & 1438800 &
0.374953 & 0.0000707557 & 4 & 606780 \\

0.374707 & 0.000337058 & 10 & 9990000 &
0.374892 & 0.000123145 & 10 & 3990000 &
0.374967 & 0.0000602778 & 10 & 2989950 \\

0.374731 & 0.000304565 & 4 & 4570380 &
0.374904 & 0.000111274 & 4 & 1438380 &
0.374975 & 0.0000544668 & 4 & 624000 \\

0.374763 & 0.000263534 & 10 & 9990000 &
0.374921 & 0.0000962831 & 10 & 3990000 &
0.374986 & 0.0000471291 & 10 & 2989950 \\

0.374782 & 0.000240375 & 4 & 4580760 &
0.374931 & 0.0000878216 & 4 & 1448760 &
0.374992 & 0.0000429874 & 4 & 621960 \\

0.374823 & 0.000193664 & 4 & 4603560 &
0.374953 & 0.0000707557 & 4 & 1443180 &
0.375005 & 0.0000346339 & 4 & 626760 \\

0.374855 & 0.000158754 & 4 & 4687200 &
0.374970 & 0.0000580011 & 4 & 1443600 &
0.375016 & 0.0000283907 & 4 & 696000 \\

\end{tabular}
\end{ruledtabular}
\end{table*}

\end{document}